\documentclass[12pt]{article}

\usepackage{amssymb,amsmath,graphics}

\numberwithin{equation}{section}

\allowdisplaybreaks

\newenvironment{Proof}{\removelastskip\par\medskip
\noindent{\em Proof.}
\rm}{\penalty-20\null\hfill$\square$\par\medbreak}

\newtheorem{prop}{Proposition}

\def\vegaa{{\mathord{{\mathrm {Vega}}}}}

\def\vannaa{{\mathord{{\mathrm {Vanna}}}}}
\def\volgaa{{\mathord{{\mathrm {Volga}}}}}
\def\vegableedaa{{\mathord{{\mathrm {Vega bleed}}}}}
\def\strikedeltaa{{\mathord{{\mathrm {Strike delta}}}}}
\def\strikegammaa{{\mathord{{\mathrm {Strike gamma}}}}}

\begin{document}
 ~
\begin{center}
{\Large Computation of second order price sensitivities in depressed markets} \\
\bigskip
{\large Youssef El-Khatib}\footnote{UAE University, Department of Mathematical Sciences, Al-Ain, P.O. Box 15551. United Arab Emirates.{\ E-mail : Youssef\_Elkhatib@uaeu.ac.ae.}
}\ \ \ \ \ {\large Abdulnasser Hatemi-J}\footnote{UAE University, Department of Economics and Finance, Al-Ain, P.O. Box 15551, United Arab Emirates. {\ E-mail : Ahatemi@uaeu.ac.ae.}}
\end{center}
~
\begin{abstract}
Risk management in financial derivative markets requires inevitably the calculation of the different price sensitivities. The literature contains an abundant amount of research works that have studied the computation of these important values. Most of these works consider the well-known Black and Scholes model where the volatility is assumed to be constant. Moreover, to our best knowledge, they compute only the first order price sensitivities. Some works that attempt to extend to markets affected by financial crisis appeared recently. However, none of these papers deal with the calculation of the price sensitivities of second order. Providing second derivatives for the underlying price sensitivities is an important issue in financial risk management because the investor can determine whether or not each source of risk is increasing at an increasing rate. In this paper, we work on the computation of second order prices sensitivities for a market under crisis. The underlying second order price sensitivities are derived explicitly. The obtained formulas are expected to improve on the accuracy of the hedging strategies during a financial crunch. 
\end{abstract}

\baselineskip=0.5cm

\baselineskip=0.5cm \noindent\textbf{Keywords:} European options, Black-Scholes model, Financial crisis, Price sensitivities, Second order price sensitivities.
\baselineskip0.7cm
\section{Introduction}
Price sensitivities are an integral part of financial risk management nowadays. A number of papers has been devoted to this issue. Current literature has provided price sensitivities for different volatility models starting with the well-known Black and Sholes (1970) model. Recently, some new ideas have been developed in order to determine whether or not each source of risk is increasing. 
Thus, within this context the computation of second order price sensitivities is a pertinent and important issue. To our best knowledge, second order price sensitivities have been introduced only for models that do not account for a market with a crisis in the existing literature. In this paper we provide second order prices sensitivities for a model of option pricing with closed form solution for a market that is characterized by a financial crisis. The second order price sensitivities that we provide are Vanna, Volga and Vega bleed, using the existing denotations from the literature. Vanna is the change in the Vega of the option with respect to the change in the asset price, which can be seen as the delta of the Vega. Volga is the second order derivative of the premium with regard to the volatility. Vega bleed is the second order of the premium with regard to a joint change in volatility and time to maturity. 
The rest of  paper is structured as follows. In Section 2 we present the option pricing for model with a crisis and we give the values of different first order price sensitivities. In section 3 we derive the different second order price sensitivities based on the suggested crisis model. The last section concludes the paper.

\section{Options pricing and price sensitivities in crisis time}
In this section, we present the crisis model, we refer the reader to \cite{savit1989}, \cite{sornette2003}, \cite{dibeh2007} for more details on modeling financial assets during crashes \footnote{ For recent papers on modeling with jump, we refer to \cite{elkhatibmdalal12} for jump-diffusion model, to \cite{elkhatibhatemi12} for price sensitivities calculation using Malliaivn calculus and to \cite{elkhatibhajji13} where a jump diffusion model during crisis is studied.}. In \cite{el2013computations}, the authors compute price sensitivities for depressed markets. In this work, we provide the second order prices sensitivities using the pricing formula obtained in \cite{elkhatibhatemi13}. \newline 
We consider a probability space $(\Omega,{\cal F},P)$ and a Brownian motion process $(W_t)_{t \in [0,T]}$ living in it. We denote by $({\cal F}_t)_{t \in [0,T]}$ the natural filtration generated by $(W_t)_{t\in [0,T]}$. The market has an European call option with underlying risky asset $S$. The return on asset without risk is denoted by $r$.
For the sake of the simplicity, we use the denotation $P$ for the risk-neutral probability. As in \cite{elkhatibhatemi13} we assume that the underlying asset price process $S=(S_t)_{t\in[0,T]}$ is governed by
 \begin{equation}
 \label{eq10}
 dS_t= r S_t dt+(\sigma S_t + \alpha e^{rt})dW_{t},
 \end{equation}
 where $t \in [0,T]$ and $S_0>0$, $\sigma$ and $\alpha$ are constant. The solution of (\ref{eq10}) is 
 \begin{equation}
 \label{eq11}
 S_t=\left(S_0+\frac{\alpha}{\sigma}\right)\exp\left[\left(r-\frac{\sigma^2}{2}\right)t+\sigma W_t\right]-\frac{\alpha}{\sigma}e^{rt}, \ \ \ t\in[0,T].
 \end{equation}
Notice that when $\alpha=0$, $S_t$ is reduced to the log-normal process of the Black-Scholes model. In the next subsection we present the option pricing formula as well as the different price sensitivities for the above crash model as derived in \cite{elkhatibhatemi13}. 
\subsection{Call-Put options prices}
The next two propositions from \cite{elkhatibhatemi13} are needed for computing the second order price sensitivities. We assume that the price process $(S_t)_{t\in [0,T]}$ under the risk-neutral probability is given by (\ref{eq11}). Let 
\begin{equation}
\label{d1}
d_1^{\alpha}= \frac{1}{\sigma \sqrt{T}}\left( \ln\left(\frac{S_0+\frac{\alpha}{\sigma}}{K+\frac{\alpha}{\sigma}e^{rT}}\right)+(r+\frac{\sigma^2}{2})T\right),
\end{equation}
and
\begin{equation}
\label{d2}
d_2^{\alpha}=  \frac{1}{\sigma \sqrt{T}}\left( \ln\left(\frac{S_0+\frac{\alpha}{\sigma}}{K+\frac{\alpha }{\sigma}e^{rT}}\right)+(r-\frac{\sigma^2}{2})T\right)=d_1^{\alpha}-\sigma\sqrt{T},
\end{equation}
and $\Phi (d)=\int_{-\infty}^d \frac{e^{-u^{2}/2}}{\sqrt{2\pi}}du.$
Then we have
\begin{prop}
The premium of an European call option with underlying asset $S=(S_t)_{t\in [0,T]}$, strike $K$ and maturity $T$ is
\begin{equation}
\label{premium1}
C(S_T, K)=E[e^{-rT}(S_T -K)^{+}]=\left(S_0+\frac{\alpha}{\sigma}\right)\Phi (d_1^{\alpha}) -\left(Ke^{-rT}+\frac{\alpha}{\sigma}\right)\Phi (d_2^{\alpha}),
\end{equation}
\end{prop}
Let $(\xi_{t,u}^x )_{u\in [t,T]}$ be the process defined as
 $$
 d\xi_{t,u}^x = r \xi_{t,u}^x du  + \sigma \xi_{t,u}^x dW_u, \ \ \ u\in
 [t,T], \ \ \ \xi_{t,t}^x = x.$$
We have $\xi_{t} = \xi_{0,t}^1$, $t\in [0,T]$.
The prices of European call and put options at any time $t$ for the crisis model is stated in the next proposition.
\begin{prop}
\label{lbk2}
The price of an European call and an European put options with underlying asset $S=(S_t)_{t\in [0,T]}$, strike $K$ and maturity $T$, at time $t\in [0,T]$, are respectively given by
$$
C(t,S_t)=\left(S_t+\frac{\alpha}{\sigma}e^{rt}\right)\Phi (d_{t,1}^{\alpha}) -\left(K e^{-r(T-t)}+\frac{\alpha}{\sigma}e^{rt}\right)\Phi (d_{t,2}^{\alpha}),
$$
and
$$
P(t,S_t)=\left(S_t+\frac{\alpha}{\sigma}e^{rt}\right) \Phi (d_{t,1}^{\alpha}) -\left(K e^{-r(T-t)}+\frac{\alpha}{\sigma}e^{rt}\right)\Phi (d_{t,2}^{\alpha})+Ke^{-r(T-t)}-S_t,
$$
where
\begin{equation}
\label{dt1}
d_{t,1}^{\alpha}= \frac{1}{\sigma \sqrt{T-t}}\left( \ln\left(\frac{S_t+\frac{\alpha}{\sigma}e^{rt}}{K+\frac{\alpha}{\sigma}e^{rT}}\right)+(r+\frac{\sigma^{2}}{2})(T-t)\right),
\end{equation}
and
$$
d_{t,2}^{\alpha}=  \frac{1}{\sigma \sqrt{T-t}}\left( \ln\left(\frac{S_t+\frac{\alpha}{\sigma}e^{rt}}{K+\frac{\alpha }{\sigma}e^{rT}}\right)+(r-\frac{\sigma^{2}}{2})(T-t)\right).
$$
\end{prop}
\subsection{Price sensitivities}
The next proposition gives the different price sensitivities for the crisis model.
\begin{prop}
	\label{gr1}
	The price sensitivities of an European call option with underlying asset $S=(S_t)_{t\in [0,T]}$, strike $K$ and maturity $T$, at time $t\in [0,T]$, are respectively given by
	\begin{eqnarray*}
		\label{delta}
		\Delta:&=&\frac{\partial C}{\partial S_t}=\Phi (d_{t,1}^{\alpha}) \\
		\label{gamma}
		\Gamma:&=&\frac{\partial^2 C}{\partial S_t^2}=\frac{e^{-(d_{t,1}^{\alpha})^{2}/2}}{(\sigma S_t +\alpha e^{rt})\sqrt{2\pi(T-t)}}=\frac{e^{-(d_{t,1}^{\alpha})^{2}/2}}{S_t\sigma\sqrt{2\pi(T-t)}}\\
		\nonumber
		\Theta:&=&\frac{\partial C}{\partial t}=-\frac{S_t \sigma+\alpha e^{rt}}{2\sqrt{2\pi \tau}}e^{-\frac{(d_{t,1}^{\alpha})^2}{2}}-rK e^{-r\tau}\Phi(d_{t,2}^{\alpha})+\frac{r\alpha}{\sigma}e^{rt}(\Phi(d_{t,1}^{\alpha})-\Phi(d_{t,2}^{\alpha}))\\
		\label{theta}
		&=&-\frac{S_t\sigma}{2\sqrt{2\pi(T-t)}}e^{-\frac{(d_{t,1}^{\alpha})^2}{2}}-rK e^{-r(T-t)}\Phi (d_{t,2}^{\alpha})+\frac{r\alpha}{\sigma}e^{rt}(\Phi(d_{t,1}^{\alpha})-\Phi(d_{t,2}^{\alpha}))\\
		\label{rho}
		\rho:&=&(T-t)K e^{-r(T-t)}\Phi (d_{t,2}^{\alpha})+\frac{\alpha t}{\sigma}e^{rt}\left(\Phi(d_{t,1}^{\alpha})-\Phi(d_{t,2}^{\alpha})\right)\\
		\label{vega}
		\nu:&=&\frac{\alpha}{\sigma^2}e^{rt}(\Phi (d_{t,2}^{\alpha})-\Phi(d_{t,1}^{\alpha}))+\frac{e^{-\frac{(d_{t,1}^{\alpha})^2}{2}}}{\sqrt{2\pi}}\left(S_t+\frac{\alpha}{\sigma}e^{rt}\right)\sqrt{T-t}.
	\end{eqnarray*}
	Moreover the price sensitivities of an European put option under similar circumstances are as follows.
	\begin{eqnarray*}
		\label{deltap}
		\Delta:&=&\Phi (d_{t,1}^{\alpha})-1 \\
		\label{gammap}
		\Gamma:&=&\frac{1}{S_t\sigma\sqrt{2\pi(T-t)}}e^{-(d_{t,1}^{\alpha})^{2}/2}\\
		\label{thetap}
		\Theta:&=&-\frac{S_t \sigma+\alpha e^{rt}}{2\sqrt{2\pi \tau}}e^{-\frac{(d_{t,1}^{\alpha})^2}{2}}-rK e^{-r\tau}\Phi(d_{t,2}^{\alpha})+\frac{r\alpha}{\sigma}e^{rt}(\Phi(d_{t,1}^{\alpha})-\Phi(d_{t,2}^{\alpha}))+r K e^{-r\tau}\\
		\label{rhop}
		\rho:&=&(T-t)K e^{-r(T-t)}\Phi (d_{t,2}^{\alpha})+\frac{\alpha t}{\sigma}e^{rt}\left(\Phi(d_{t,1}^{\alpha})-\Phi(d_{t,2}^{\alpha})\right)-\tau K e^{-r\tau}\\
		\label{vegap}
		\nu:&=&\frac{\alpha}{\sigma^2}e^{rt}(\Phi (d_{t,2}^{\alpha})-\Phi(d_{t,1}^{\alpha}))+\frac{e^{-\frac{(d_{t,1}^{\alpha})^2}{2}}}{\sqrt{2\pi}}\left(S_t+\frac{\alpha}{\sigma}e^{rt}\right)\sqrt{T-t}.
	\end{eqnarray*}
\end{prop}
For the proof of the above proposition see \cite{elkhatibhatemi13}. 
\section{Calculation of second order price sensitivities in crisis times}
In this section, we compute second order price sensitivities for the crisis model. More precisely, we compute the following second order derivatives: 
\begin{eqnarray*}
\vannaa:&=& \frac{\partial \vegaa}{\partial S}=\frac{\partial^2 C}{\partial S \partial \sigma},\ \ \ \ \ \ \volgaa:=\frac{\partial \vegaa}{\partial \sigma} =\frac{\partial^2 C}{\partial \sigma^2},\\
\vegableedaa:&=& \frac{\partial^2 C}{\partial T \partial \sigma},\ \ \ \ \ \ \strikegammaa:=\frac{\partial \strikedeltaa}{\partial K}=\frac{\partial^2 C}{\partial K^2}.
\end{eqnarray*}
\begin{prop}
	\label{sgr1}
The second order sensitivities of an European call option with underlying asset $S=(S_t)_{t\in [0,T]}$, strike $K$ and maturity $T$, at time $t\in [0,T]$, are respectively given by
\begin{eqnarray*}
		\label{vannaformula}
		\vannaa &=&\frac{\partial^2 C}{\partial S \partial \sigma}= \frac{\alpha }{\sigma^2} \frac{S_t-K e^{-r\tau}}{K e^{-rT}+\frac{\alpha}{\sigma}}	\Gamma -\frac{d_{t,2}^{\alpha}e^{-\frac{(d_{t,1}^{\alpha})^2}{2}}}{\sigma\sqrt{2\pi}}.\\
		\label{vomma}
		\volgaa &=&\frac{\partial^2 C}{\partial \sigma^2}=\frac{-2\alpha }{\sigma^3} e^{rt}\left(\Phi (d_{t,2}^{\alpha})-\Phi(d_{t,1}^{\alpha})\right)\\
		&&+\frac{\alpha }{\sigma^2} e^{rt}\left(\frac{e^{-\frac{(d_{t,2}^{\alpha})^{2}}{2}}}{\sqrt{2\pi}}\frac{\partial d_{t,2}^{\alpha}}{\partial \sigma}-\frac{e^{-\frac{(d_{t,1}^{\alpha})^{2}}{2}}}{\sqrt{2\pi}}\frac{\partial d_{t,1}^{\alpha}}{\partial \sigma}\right)\\
		&&+\frac{\sqrt{\tau}}{\sqrt{2\pi}}e^{-\frac{(d_{t,1}^{\alpha})^2}{2}}\left[
		1-d_{t,1}^{\alpha} \left(S_t+\frac{\alpha}{\sigma}e^{rt}\right)
		\frac{\partial  d_{t,1}^{\alpha}}{\partial \sigma} \right].\\
		\label{vegableedaa}
		\vegableedaa&=&\frac{\partial^2 C}{\partial T \partial \sigma}=\frac{\alpha }{\sigma^2} e^{rt} \left(\frac{e^{-\frac{(d_{t,2}^{\alpha})^{2}}{2}}}{\sqrt{2\pi}}\frac{\partial d_{t,2}^{\alpha}}{\partial \tau}-\frac{e^{-\frac{(d_{t,1}^{\alpha})^{2}}{2}}}{\sqrt{2\pi}}\frac{\partial d_{t,1}^{\alpha}}{\partial \tau}\right)\\
		&&+\frac{\sqrt{\tau}}{\sqrt{2\pi}}e^{-\frac{(d_{t,1}^{\alpha})^2}{2}}\left(S_t+\frac{\alpha}{\sigma}e^{rt}\right)\left[
		\frac{1}{2\tau}-d_{t,1}^{\alpha} 
		\frac{\partial  d_{t,1}^{\alpha}}{\partial \tau} \right].\\
		\label{strikegammaformula}
		\strikegammaa &=&\frac{\partial^2 C}{\partial K^2}.
\end{eqnarray*}
Where
\begin{eqnarray}
\label{partial d1sigma}
\frac{\partial d_{t,2}^{\alpha}}{\partial \sigma}&=&\frac{-d_{t,1}}{\sigma}+\frac{\alpha}{\sigma^3 \sqrt{\tau}}e^{rt}\left[\frac{1}{K e^{-r\tau}+\frac{\alpha}{\sigma}e^{rt}}\frac{S_t-K e^{-r\tau}}{S_t+\frac{\alpha}{\sigma}e^{rt}}\right],\\
\label{partial d2sigma}
\frac{\partial d_{t,2}^{\alpha}}{\partial \sigma}&=&\frac{\partial d_{t,1}^{\alpha}}{\partial \sigma}-\sqrt{\tau}.
\end{eqnarray} 
And
$$
\frac{\partial d_{t,2}^{\alpha}}{\partial \tau}=\frac{\partial d_{t,1}^{\alpha}}{\partial \tau}-\frac{\sigma}{2\sqrt{\tau}},
$$ 
where
\begin{equation}
\label{partial d1tau}
\frac{\partial d_{t,1}^{\alpha}}{\partial \tau}= \frac{1}{2\sigma \tau \sqrt{\tau}}\left( \ln\left(\frac{S_t+\frac{\alpha}{\sigma}e^{rt}}{K+\frac{\alpha}{\sigma}e^{rT}}\right)
+(r+\frac{\sigma^{2}}{2})\tau\right)+\frac{1}{2\sigma \tau \sqrt{\tau}}\left(r+\frac{\sigma^{2}}{2}\right).
\end{equation}
\end{prop}
\begin{Proof}
We make use of the first order price sensitivities stated in proposition~\ref{gr1}. 
Then $\vannaa$ can be computed by differentiating $\vegaa$ with respect to $S$ the underlying asset price,  as follows
\begin{eqnarray*}
	\vannaa &=&\frac{\partial^2 C}{\partial S \partial \sigma}= \frac{\partial \nu}{\partial S_t}= \frac{\partial \left[ \frac{\alpha}{\sigma^2}e^{rt}(\Phi (d_{t,2}^{\alpha})-\Phi(d_{t,1}^{\alpha}))+\frac{e^{-\frac{(d_{t,1}^{\alpha})^2}{2}}}{\sqrt{2\pi}}\left(S_t+\frac{\alpha}{\sigma}e^{rt}\right)\sqrt{T-t}\right]}{\partial S_t}\\		
		&=& \frac{\alpha }{\sigma^2} e^{rt}\frac{\partial  }{\partial S_t} \left(\Phi (d_{t,2}^{\alpha})-\Phi(d_{t,1}^{\alpha})\right)+\frac{\sqrt{T-t}}{\sqrt{2\pi}}e^{-\frac{(d_{t,1}^{\alpha})^2}{2}}\left[
		1-d_{t,1}^{\alpha} \left(S_t+\frac{\alpha}{\sigma}e^{rt}\right)
		\frac{\partial  d_{t,1}^{\alpha}}{\partial S_t} \right].	
\end{eqnarray*}
But notice that
$$\frac{\partial \Phi (d_{t,1}^{\alpha})}{\partial S_t}=\Gamma \ \ \ \mbox{and} \ \ \ \frac{\partial d_{t,1}^{\alpha} }{\partial S_t}=\frac{1}{\sigma \sqrt{\tau}(S_t +\frac{\alpha}{\sigma}e^{rt})}$$
\begin{eqnarray*}
\frac{\partial \Phi(d_{t,2}^{\alpha})}{\partial S_t}&=&\frac{\partial \Phi(d_{t,2}^{\alpha})}{\partial d_{t,2}^{\alpha}}\frac{\partial d_{t,2}^{\alpha} }{\partial d_{t,1}^{\alpha}}\frac{\partial d_{t,1}^{\alpha} }{\partial S_t}=\frac{\partial \Phi(d_{t,2}^{\alpha})}{\partial d_{t,2}^{\alpha}}\frac{\partial d_{t,1}^{\alpha} }{\partial S_t}\\
&=& \frac{e^{-\frac{(d_{t,1}^{\alpha}-\sigma \sqrt{\tau})^{2}}{2}}}{\sqrt{2\pi}}\frac{\partial d_{t,1}^{\alpha} }{\partial S_t}=\frac{e^{-\frac{(d_{t,1}^{\alpha})^{2}}{2}-\frac{\sigma^{2}\tau}{2}+ d_{t,1}^{\alpha}\sigma\sqrt{\tau}}}{\sqrt{2\pi}}\frac{\partial d_{t,1}^{\alpha} }{\partial S_t}\\
&=&\frac{\partial \Phi(d_{t,1}^{\alpha})}{\partial d_{t,1}^{\alpha}}\frac{\partial d_{t,1}^{\alpha} }{\partial S_t}\frac{S_t+\frac{\alpha}{\sigma}e^{rt}}{K e^{-r\tau}+\frac{\alpha}{\sigma}e^{rt}}=\frac{\partial \Phi(d_{t,1}^{\alpha})}{\partial S_t}\frac{S_t+\frac{\alpha}{\sigma}e^{rt}}{K e^{-r\tau}+\frac{\alpha}{\sigma}e^{rt}}\\
&=&\Gamma \frac{S_t+\frac{\alpha}{\sigma}e^{rt}}{K e^{-r\tau}+\frac{\alpha}{\sigma}e^{rt}},
\end{eqnarray*}
where $\tau:=T-t$ is the time to maturity. Thus
\begin{eqnarray*}
	\vannaa &=&\frac{\partial^2 C}{\partial S \partial \sigma}= \frac{\partial \nu}{\partial S_t}= \frac{\partial \left[ \frac{\alpha}{\sigma^2}e^{rt}(\Phi (d_{t,2}^{\alpha})-\Phi(d_{t,1}^{\alpha}))+\frac{e^{-\frac{(d_{t,1}^{\alpha})^2}{2}}}{\sqrt{2\pi}}\left(S_t+\frac{\alpha}{\sigma}e^{rt}\right)\sqrt{T-t}\right]}{\partial S_t}\\		
	&=& \frac{\alpha }{\sigma^2} e^{rt} \Gamma \frac{S_t-K e^{-r\tau}}{K e^{-r\tau}+\frac{\alpha}{\sigma}e^{rt}}	+\frac{\sqrt{\tau}}{\sqrt{2\pi}}e^{-\frac{(d_{t,1}^{\alpha})^2}{2}}\left[
	1-d_{t,1}^{\alpha} \left(S_t+\frac{\alpha}{\sigma}e^{rt}\right)
	\frac{1}{\sigma \sqrt{\tau}(S_t +\frac{\alpha}{\sigma}e^{rt})}\right]\\
	&=&\frac{\alpha }{\sigma^2} e^{rt} \Gamma \frac{S_t-K e^{-r\tau}}{K e^{-r\tau}+\frac{\alpha}{\sigma}e^{rt}}	+\frac{e^{-\frac{(d_{t,1}^{\alpha})^2}{2}}}{\sigma\sqrt{2\pi}}(
	\sigma \sqrt{\tau}-d_{t,1}^{\alpha} ),
	\end{eqnarray*}
	which gives (\ref{vannaformula}). Similarly, $\volgaa$ can be computed as follows:
\begin{eqnarray*}
	\volgaa &=&\frac{\partial^2 C}{\partial^2 \sigma}= \frac{\partial \nu}{\partial \sigma}= \frac{\partial \left[ \frac{\alpha}{\sigma^2}e^{rt}(\Phi (d_{t,2}^{\alpha})-\Phi(d_{t,1}^{\alpha}))+\frac{e^{-\frac{(d_{t,1}^{\alpha})^2}{2}}}{\sqrt{2\pi}}\left(S_t+\frac{\alpha}{\sigma}e^{rt}\right)\sqrt{T-t}\right]}{\partial \sigma}\\		
	&=& \frac{-2\alpha }{\sigma^3} e^{rt}\left(\Phi (d_{t,2}^{\alpha})-\Phi(d_{t,1}^{\alpha})\right)+\frac{\alpha }{\sigma^2} e^{rt}\frac{\partial  }{\partial \sigma} \left(\Phi (d_{t,2}^{\alpha})-\Phi(d_{t,1}^{\alpha})\right)\\
	&&+\frac{\sqrt{\tau}}{\sqrt{2\pi}}e^{-\frac{(d_{t,1}^{\alpha})^2}{2}}\left[
	1-d_{t,1}^{\alpha} \left(S_t+\frac{\alpha}{\sigma}e^{rt}\right)
	\frac{\partial  d_{t,1}^{\alpha}}{\partial \sigma} \right],
\end{eqnarray*}
with
\begin{eqnarray*}
\frac{\partial  }{\partial \sigma} \left(\Phi (d_{t,2}^{\alpha})-\Phi(d_{t,1}^{\alpha})\right)&=&\frac{\partial \Phi (d_{t,2}^{\alpha}) }{\partial \sigma} -\frac{\partial  \Phi (d_{t,1}^{\alpha})}{\partial \sigma} \\
&=&\frac{\partial \Phi(d_{t,2}^{\alpha})}{\partial d_{t,2}^{\alpha}}\frac{\partial d_{t,2}^{\alpha}}{\partial \sigma}-\frac{\partial \Phi(d_{t,1}^{\alpha})}{\partial d_{t,1}^{\alpha}}\frac{\partial d_{t,1}^{\alpha}}{\partial \sigma}\\
&=&\frac{e^{-\frac{(d_{t,2}^{\alpha})^{2}}{2}}}{\sqrt{2\pi}}\frac{\partial d_{t,2}^{\alpha}}{\partial \sigma}-\frac{e^{-\frac{(d_{t,1}^{\alpha})^{2}}{2}}}{\sqrt{2\pi}}\frac{\partial d_{t,1}^{\alpha}}{\partial \sigma},
\end{eqnarray*}
with
\begin{eqnarray*}
\frac{\partial d_{t,2}^{\alpha}}{\partial \sigma}&=&\frac{-d_{t,1}}{\sigma}+\frac{\alpha}{\sigma^3 \sqrt{\tau}}e^{rt}\left[\frac{1}{K e^{-r\tau}+\frac{\alpha}{\sigma}e^{rt}}\frac{S_t-K e^{-r\tau}}{S_t+\frac{\alpha}{\sigma}e^{rt}}\right],\\
\frac{\partial d_{t,2}^{\alpha}}{\partial \sigma}&=&\frac{\partial d_{t,1}^{\alpha}}{\partial \sigma}-\sqrt{\tau}.
\end{eqnarray*} 
The Vega Bleed is the change of the Vega when there is a time change. We calculate the Vega Bleed as follows:
\begin{eqnarray*}
	\vegableedaa &=&\frac{\partial^2 C}{\partial \tau \partial \sigma}= \frac{\partial \nu}{\partial \tau}= \frac{\partial \left[ \frac{\alpha}{\sigma^2}e^{rt}(\Phi (d_{t,2}^{\alpha})-\Phi(d_{t,1}^{\alpha}))+\frac{e^{-\frac{(d_{t,1}^{\alpha})^2}{2}}}{\sqrt{2\pi}}\left(S_t+\frac{\alpha}{\sigma}e^{rt}\right)\sqrt{\tau}\right]}{\partial \tau}\\		
	&=& \frac{\alpha }{\sigma^2} e^{rt} \left(\frac{e^{-\frac{(d_{t,2}^{\alpha})^{2}}{2}}}{\sqrt{2\pi}}\frac{\partial d_{t,2}^{\alpha}}{\partial \tau}-\frac{e^{-\frac{(d_{t,1}^{\alpha})^{2}}{2}}}{\sqrt{2\pi}}\frac{\partial d_{t,1}^{\alpha}}{\partial \tau}\right)\\
	&&+\frac{\sqrt{\tau}}{\sqrt{2\pi}}e^{-\frac{(d_{t,1}^{\alpha})^2}{2}}\left(S_t+\frac{\alpha}{\sigma}e^{rt}\right)\left[
	\frac{1}{2\tau}-d_{t,1}^{\alpha} 
	\frac{\partial  d_{t,1}^{\alpha}}{\partial \tau} \right].	
\end{eqnarray*}
Notice that
$$
\frac{\partial d_{t,2}^{\alpha}}{\partial \tau}=\frac{\partial d_{t,1}^{\alpha}}{\partial \tau}-\frac{\sigma}{2\sqrt{\tau}}
$$
and
$$
\frac{\partial d_{t,1}^{\alpha}}{\partial \tau}=  \frac{1}{2\sigma \tau \sqrt{\tau}}\left( \ln\left(\frac{S_t+\frac{\alpha}{\sigma}e^{rt}}{K+\frac{\alpha}{\sigma}e^{rT}}\right)
+(r+\frac{\sigma^{2}}{2})\tau\right)+\frac{1}{2\sigma \tau \sqrt{\tau}}(r+\frac{\sigma^{2}}{2}),
$$
which ends the proof.
\end{Proof}

\section{Conclusions}
Price sensitivities are regularly used by financial institutions and investors in order to deal with different sources of financial risk. Recently, the literature has put forward formulas for price sensitivities for markets that are characterized by a crisis. This is an important issue because it is exactly during the crisis that the need for successful tools that can neutralize or at least reduce risk is urgent. Another strand of literature has contributed to the introduction of the second order price sensitivities.  To the best knowledge, the second order price sensitivities have not been developed for a market with a crisis. Thus, our main goal in this paper is to introduce the second order price sensitivities for a market that is potentially experiencing a crisis. All the new suggested formulas are expressed as propositions and each underlying proposition is provided with a mathematical proof. The formulas that are proposed in this paper are expected to make the hedging against the financial risk more precise.

\end{document}